\documentclass[12pt,preprint]{aastex}
\usepackage{natbib}

\shorttitle{Methane cells for precision RVs in the K band}
\shortauthors{Anglada-Escud\'e, Plavchan, et al.}

\begin{document}

\title{Design and Construction of Absorption Cells for Precision Radial
Velocities in the K Band using Methane Isotopologues}

\author{
Guillem Anglada-Escud\'e\altaffilmark{1}, 
Peter Plavchan\altaffilmark{2},
Sean Mills\altaffilmark{3},
Peter Gao\altaffilmark{3},
Edgardo Garc\'ia-Berr\'ios\altaffilmark{4},
Nathan S. Lewis\altaffilmark{4},
Keeyoon Sung\altaffilmark{5},
David Ciardi\altaffilmark{2},
Chas Beichman\altaffilmark{2},
Carolyn Brinkworth\altaffilmark{2},
John Johnson\altaffilmark{3},
Cassy Davison\altaffilmark{6}, 
Russel White\altaffilmark{6}, 
Lisa Prato\altaffilmark{7}.
}

\email{
anglada@dtm.ciw.edu, 
plavchan@ipac.caltech.edu}

\altaffiltext{1}{Carnegie Institution of Washington, Department of Terrestrial 
Magnetism, 5241 Broad Branch Rd. NW, Washington D.C., 20015, USA}
\altaffiltext{2}{NASA Exoplanet Science Institute, 770 S Wilson Ave, M/C 100-22, Pasadena, CA 91125, USA}
\altaffiltext{3}{Caltech, , MC 249-17, 1200 East California Blvd, Pasadena CA 91125 USA}
\altaffiltext{4}{Noyes Laboratory 210, 127-72, Division of Chemistry and Chemical Engineering, 
California Institute of Technology, Pasadena, California 91125}
\altaffiltext{5}{Jet Propulsion Laboratory, California Institute of
Technology, 4800 Oak Grove Dr, Pasadena, CA 91109, USA}
\altaffiltext{6}{Georgia State University, Department of Physics 
and Astronomy, 29 Peachtree Center Avenue, Science Annex, Suite 400, 
Atlanta, GA 30303}
\altaffiltext{7}{Lowell Observatory,
1400 West Mars Hill Road. Flagstaff, AZ 86001
}

\begin{abstract}  

We present a method to optimize absorption cells for precise
wavelength calibration in the near-infrared. We apply it to design
and optimize methane isotopologue cells for precision radial velocity
measurements in the K band. We also describe the construction and
installation of two such cells for the CSHELL spectrograph at
NASA's IRTF. We have obtained their high-resolution laboratory
spectra, which we can then use in precision radial velocity measurements
and which can also have other applications. In terms of obtainable
RV precision methane should out-perform other proposed cells, such as
the ammonia cell ($^{14}$NH$_{3}$) recently demonstrated on
CRIRES/VLT. The laboratory spectra of Ammonia and the Methane cells
show strong absorption features in the H band that could also be
exploited for precision Doppler measurements. We present spectra
and preliminary radial velocity measurements obtained during our
first-light run. These initial results show that a precision down
to 20-30 m s$^{-1}$ can be obtained using a wavelength interval of only 5
nm in the K band and S/N$\sim$150. This supports the prediction that
a precision down to a few m s$^{-1}$ can be achieved on late M dwarfs
using the new generation of NIR spectrographs, thus enabling the
detection of terrestrial planets in their habitable zones. Doppler
measurements in the NIR can also be used to mitigate the radial
velocity jitter due to stellar activity enabling more efficient
surveys on young active stars.

\end{abstract}


\section{Introduction} 

The radial velocity (RV) technique is the most efficient method to
detect planetary-mass bodies orbiting stars. Accuracies at the
level of $\sim$1 m s$^{-1}$ have been recently achieved with
state-of-the-art optical spectrographs such as HARPS/ESO
\citep[]{mayor:2009}, HIRES/Keck \citep[]{howard:2010} or
PFS/Magellan \citep[]{crane:2010}. The radial velocity technique
is most sensitive to massive planets around low-mass stars, but
such stars are intrinsically faint at optical wavelengths and only
a handful of nearby and relatively massive M dwarfs have been
successfully monitored for planetary
systems \citep[e.g.,][]{endl:2006, johnson:2007, zec:2009}.
Although the number of M dwarfs surveyed in the optical is small,
they have produced some of the most  spectacular results in the
field: multiplanetary systems with several super-Earths \citep[GJ
581 : ][]{mayor:2009,vogt:2010}, the first transiting Neptune mass
planets \citep[GJ 436, GJ 1214, ][]{gillon:2007,charbonneau:2009},
and the most dynamically complex systems with both giant planets
and super-Earth-mass bodies \citep[GJ 876bcde : ][]{rivera:2010}.
Thus, there is strong statistical evidence that M-dwarfs are rich
in sub-Neptune-mass planets \citep{mayor:2009} and (possibly)
Earth-mass planets as well \citep[]{howard:2010}.

Since most of the flux of M dwarfs is emitted in the near-infrared
(NIR), many more and later-type M dwarfs could be surveyed
provided that adequate wavelength-calibration techniques and
spectrographs are developed in the NIR region \citep[e.g.,
][]{reiners:2010}. Additionally, young and/or active stars will
have relatively more quiescent photospheres in the nIR relative to
the optical, allowing for wavelength-dependent characterization or
mitigation of stellar jitter \citep{bailey:2011}  .

Recently, \citet[]{bean:2010a} have shown that accuracy at the
$\sim$5-10 m s$^{-1}$ level can be achieved on timescales of several
months using an absorption cell filled with ammonia gas ($^{14}$NH$_{3}$). 
This ammonia cell has been used to rule out the presence of the
astrometric planet candidate around the low-mass M8.5V star VB10
\citep[]{bean:2010b}. Fostered by the success of this pioneering
technique, we started a collaboration to design, build, and
implement optimized absorption cells on the available (and near
future) NIR spectrographs.

During our investigation, we found that methane is, in fact, a
very suitable gas for wavelength calibration in the K band. The
frequency precision and accuracy of CH$_4$ achieved by
\citet{titan:2009} \footnote{ Methane on Titan project,
mantained by Vincent Boudon,
\texttt{http://www.icb.cnrs.fr/titan/}}
indicates that methane absorption features can be good enough 
for RV observations. The viability of $^{12}$CH$_{4}$ as a
wavelength standard at other wavelengths is well proven. We
refer the reader to the \textit{Bureau International des Poids
et
Mesures}\footnote{\texttt{http://www.bipm.org/en/publications/mep.html}},
or \citet{methaneref1} for examples detailing the use of
methane as a wavelength-calibration standard in other lines of
inquiry. Despite this, it is well known to astronomers
working in the near infrared domain (1.0 to 5 microns) that the
Earth's atmosphere contains sufficient methane to also produce
deep absorption features, especially beyond 2.0 $\mu$m.
Therefore, it would be difficult to accurately disentangle
these telluric absorption features from those created by a cell
containing standard $^{12}$CH$_{4}$. However, carbon and
hydrogen have other stable isotopes that also form chemically
stable isotopologues of methane. The substitution of an atom by
another isotope significantly shifts the absorption features of
a molecule. As shown later, we find that such a shift is large
enough to avoid confusion of the methane isotopologues with the
more common $^{12}$CH$_{4}$. The preliminary design of optimal
gas cells is done using the line lists available in the
HITRAN 2008 database \citep{hitran:2008}. We concentrate this
study on the two simpler methane isotopologues $^{13}$CH$_{4}$
(methane-13) and $^{12}$CH$_{3}$D (deuterated methane) and
compare their performance to ammonia ($^{14}$NH$_{3}$). Since we
have built a science quality cell for each gas, we also provide the
construction details for such cells. 

%We obtained FTIR
%spectra of all the cells from 1.0 to 5.0 microns, and we also
%show their absorption spectrum in the H band. A specific cell
%optimized for the H band would be potentially useful to study
%young active early type stars (G and K dwarfs), but we don't do
%this optimization here for the sake of brevity.

The $^{13}$CH$_{4}$ cell has been installed and used in a prototype
program on the CSHELL spectrograph at NASA's Infrared Telescope
Facility (IRTF). We present the first-light spectra of bright stars
through this cell, illustrating that the proposed setup is ready to
begin precision RV measurements. The use of the cell and the Fourier
trasform infrared spectra (FTIR) are available to the community.

\section{Optimizing a Gas Cell for a Spectrograph}\label{sec:cons}

\subsection{Free Parameters}

When using a gas absorption cell for precise wavelength calibration,
its transmission absorption spectrum will determine the maximum
achievable RV precision. The cell absorption spectra mainly depend on
the following parameters: length of the cell, gas used, gas
temperature, gas pressure and spectral resolution set by the
spectrograph. Among these parameters, the only relevant freedom is
the choice of the gas and its pressure. For practical reasons, the
cell temperature should be around 300 K. A few tens of Kelvins do not
make a substantial difference in the absorption spectra of the gases
under consideration. Longer optical paths produce deeper and sharper
features, which are both desirable to obtain a better
wavelength-calibration setup.  Therefore, the cell length should be
as long as physically allowed by the spectrograph \citep[e.g. $\sim$
20 cm : ][]{bean:2010a}. A cell with multiple reflections could be
used to increase the optical depth at the cost of a more bulky setup
and some losses in each reflection \citep[see][as an
example]{mahadevan:2009}. We will not discuss this option here. 

The spectral resolution is a measure of the smallest separation
$\delta \lambda$ at which spectral features can be distinguished.
In astronomical spectrographs, it is usually defined in relation to
the resolving power $R$=$\lambda/\delta \lambda$ which is ideally
constant with wavelength. To obtain the maximal RV precision,
$\delta \lambda$ needs to be as small as possible or, equivalently,
$R$ has to be as high as possible. The stellar spectral features
have to be resolved when using traditional spectroscopy to measure
precise RVs \citep[as opposed to externally dispersed
interferometers, see ][as an example]{ge:2002}. The range of
available spectral resolution for precise RV measurements is around
R$\sim$30~000 \citep[e.g. NIRSPEC/Keck, ][]{nirspec:1998} to
$R$=110~000 \citep[CRIRES/VLT, ][]{crires:2004}. As an example,
$R$=30~000 implies that at $2300$ nm one can resolve a $\delta
\lambda$ of $0.076$ nm, while $0.0209$ nm can be resolved if $R$ =
110~000. As shown later, the resolution of the spectrograph is a
critical element in the choice of the right gas and pressure. As a
general rule, $\delta \lambda$ is defined as the
full width at half maximum (FWHM) of the point spread function (PSF)
in the wavelength dispersion direction. This PSF (also called
instrumental profile) is intrinsic to each instrument and has
nothing to do with the physical processes involved in the
generation of absorption lines in the intervening gas or the
stellar spectrum. For simulation purposes, the shape of this
instrumental profile can be approximated by a Gaussian profile or an
ensemble of Gaussian profiles.  The precise shape will only be
relevant in the actual reduction of the observations and will be
different for each instrument. Strictly speaking, a Gaussian
profile has a $\sigma = FWHM/2.35$, however typical instrumental
profiles tend to have higher wings effectively degrading the actual
resolution. For the purpose of quantifying the dependence of the
maximum precision as a function of the spectral resolution, we
assume tha the instrumental profile is a Gaussian with 
$\sigma=\delta \lambda/2$. The product of the stellar spectrum, 
the absorption spectrum of the
gas cell, and the absorption of the atmosphere have to be convolved
with this instrumental profile to obtain the observed (simulated)
spectrum.

In summary, given room-temperature operating conditions of
$\sim$300 K, a cell length of the order of $\sim$10 cm, and an
optimal spectral resolution (depends on the spectrograph design
details), the gas pressure is the only free parameter to adjust to
reach the maximal RV precision.

\subsection{Choice of Gases and Optimization Metric}

In the preliminary phase of our investigation, we were interested
in assessing which gas was more adequate for RV measurements in
the K band. To quantitatively compare the nominal performance of a
gas cell paired with a spectrograph, we used the photon-noise-limited 
precision $\sigma_{V}$ as derived by \citet{butler:1996}
as our metric. The photon-noise-limited precision has two
components: the contribution of the gas cell $\sigma_{c}$ and the
contribution of the stellar spectrum $\sigma_{*}$. The
contribution of the gas cell $\sigma_{c}$ represents how well the 
wavelength of each pixel can be measured, while $\sigma_*$
represents how well a stellar Doppler offset can be measured given the
richness of spectral features on the stellar spectrum.
The expression for $\sigma_{V}$ reads 

\begin{eqnarray}
\sigma_{V} &=& \sqrt{\sigma^2_{c} + \sigma^2_{*}}\, , \\
\sigma_{c} &=& 
  c\left(\sum_{pix} 
  \lambda\frac{dI_{c}}{d\lambda} \times {\rm S/N}\right)^{-1/2}\, , \\
\sigma_{*} &=& 
  c\left(\sum_{pix} 
  \lambda\frac{dI_*}{d\lambda} \times {\rm S/N}\right)^{-1/2} \, .  \nonumber \\
\end{eqnarray}

\noindent where $c$, when not a subscript, is the speed of light;
$I$ is the intensity of the stellar spectrum ($*$) or the cell ($c$)
spectra normalized to a continum equal to 1; $\lambda$ is the
wavelength in meters; and S/N is the signal-to-noise ratio at each
element of the sum and equals to $\sqrt{N_{\rm{phot}}}$ assuming
Poisson statistics. The sum is calculated over all the resolution
elements $\delta \lambda$. Usually, modern spectrographs are
designed in such a way that each $\delta \lambda$ is covered by 2 or
more sampling elements (or pixels). As long as there is more than
one pixel on each $\delta \lambda$, the number of pixels used does
not affect the nominal photon-noise limit. For example, lets assume
a S/N of 100 on each $\delta \lambda$ (N$_{\rm{photons}}$ = 10 000).
If we have 2 pixels on each $\delta \lambda$, each pixel will
collect 5000 photons and a corresponding S/N per pixel of 70.7. However,
the loss of S/N per pixel is compensated in equation (1) by having twice
the number of elements in the sum. In reality, a larger number of
pixels (e.g., $>2$) per $\delta \lambda$ is always better to model
the instrumental profile. The sampling of the profile will contribute to the
final error budget irrespective of the chosen calibration gas so, in
a relative sense, it does not affect our comparison metric. 

The photon-noise-limited precision $\sigma_{V}$ is the function to be minimized
with respect to the gas parameters (as we discussed, only pressure). Note that
this is an ideal estimate of the final RV precision. Real observations will
contains additional sources of uncertainty, such as the detector performance,
instrumental profile modelling, availability of adequate stellar templates, and
contamination by telluric features. At this time, the major limitation to
achieve high precision is the limited number of high-resolution spectrographs
with near-infrared capabilities and the limited wavelength range they can
provide in each single exposure. The only instrument able to deliver spectral
resolution over $10^5$ is CRIRES/VLT, and it is still a single-order
spectrograph covering only 40 nm at 2.3 $\mu$s \citep{bean:2010a}. The quality
and size of the NIR imaging arrays is also a limiting factor on some instruments
such as NIRSPEC/Keck \citep{bailey:2011}, which is also limited by a relatively
low resolution $R\sim30 000$. It is expected that new NIR spectrographs will
have a significantly increased spectral grasp and will incorporate newly
developed higher-quality CMOS imaging arrays\footnote{e.g.
\texttt{http://www.teledyne-si.com/infrared\_sensors/index.html}}, greatly
mitigating the systematic uncertainties due to the detector performance on
fainter targets.

\subsection{Methane versus ammonia}

The two gases we compare here are ammonia and methane.  \citet{nh3:2005,
nh3:2008} predicted that the NIR spectral features of ammonia were useful as a
frequency-calibration source for RV observations \citep[see also][]{urban:1989}.
A cell with the most common isotopologue of ammonia (from now on
$^{14}$NH$_{3}$) has already been demonstrated at the telescope
\citep{bean:2010a, bean:2010b}, so we wanted to assess if it was worth
developing a brand new cell based on an alternative gas (Methane). Both gases
good-quality  line lists in the HITRAN 2008 database  \citep[beyond 1.9 microns
for Ammonia and from 1.0 to $>5$ microns for Methane, ][]{hitran:2008}, allowing
straighforward simulations of the spectra for a given set of cell parameters.
Ammonia also has abundant spectral features below 1.6 $\mu$m and they have been
recently reported by \citet{sung:2012}, but at the time we build our cells,
those lines were unknown to us and we do not discuss them in detail here. Some
alternative gases for work in the H band have been given by
\citet{valdivielso:2010} and \citet{mahadevan:2009}. The method of cell
optimization presented herein can be applied to any other gas given the required
information to generate synthetic absorption spectra (line lists or public FTIR
spectra). As a matter of fact, a simplified version of the analysis presented
here was done on most of the gases available in the HITRAN 2008 database.
Methane was identified as a promising gas during such quick-look analysis. 

\begin{figure}[tb]
  \begin{center}
    \includegraphics[width=6.05in, clip]{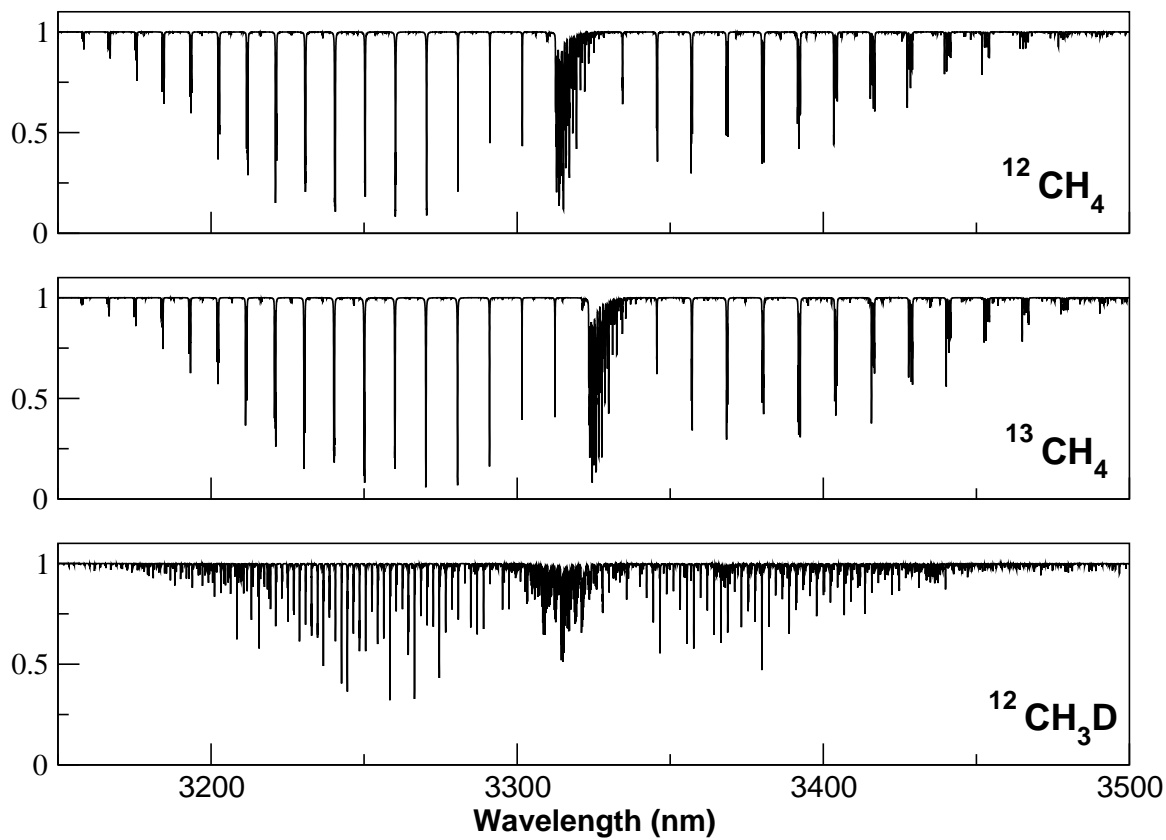}
  \end{center}
  
  \caption{Synthetic absorption at 3.0 microns of the 3
  isotopologues of methane available at the HITRAN 2008 database.
  Methane is a very strong absorber at this wavelength, so cell
  lengths of a few mm or cm have been used to generate this
  example spectra (T=300 K, P=300 mb). 
  }
  \label{fig:3microns}
\end{figure}

As discussed before, it is well known that telluric methane
features are omnipresent in NIR spectra (especially in the K and
redder bands). In order to avoid blends of the calibration spectrum
with telluric features, we propose using methane isotopologues
instead ($^{13}$CH$_{4}$ and $^{12}$CH$_{3}$D). At the obtained
working pressures and temperatures, they are both easy to handle
and much less reactive than $^{14}$NH$_{3}$. It also helps that
enough gas to build several cells could be purchased for a few
hundred US dollars. The NIR absorption features of methane are
molecular rotovibrational transitions related to the C-H bond and
the moment of inetia the methane molecule. The main difference
between $^{12}$CH$_{4}$ and $^{13}$CH$_{4}$ is a change in the
reduced mass of the $^{13}$C-H bond, changing the wavelengths of
the $^{12}$CH$_{4}$ transitions by a multiplicative factor. In the
case of $^{12}$CH$_{3}$D, the substitution of a C-H by a C-D bond
adds an additional oscillator and breaks the tetrahedric symmetry
of the molecule. As a result the $^{12}$CH$_{3}$D spectrum is a
\textit{scrambled} version of $^{12}$CH$_{4}$. Even though there is
extensive literature on the interpretation of the absorption
spectrum for both isotopologues, no comprehensive line lists are
readily available in a straightforward format. The HITRAN 2008
database contains some lines of all three methane isotopologues in
a narrow range between 3.0 and 3.5 $\mu$m (see Fig.
\ref{fig:3microns}). Using the centers of the sharper lines between
3.0 and 3.5 $\mu$m, we find that a multiplicative factor of
$\sim$1.0032 in needed to reproduce the wavelength shift in the
lines of $^{13}$CH$_{4}$ with respect to $^{12}$CH$_{4}$.  This
number is a simplification of the complex rovibrational spectral
transition changes between the two isotopes.  However, this
multiplicative factor approximation enables one to obtain a
realistic spectrum of $^{13}$CH$_{4}$ in terms of the approximate
line density and depth as a function of wavelength and to
consequently evaluate the performance of $^{13}$CH$_{4}$ as a
wavelength-calibration gas. At the K band ($\lambda\sim$ 2300 nm),
this translates into a shift of $\sim 8$ nm with respect the
equivalent features in $^{12}$CH$_{4}$. More importantly, this
shift is more than sufficient to avoid blends with telluric
$^{12}$CH$_{4}$ features (typical width of 0.1 nm at the K band,
see Section \ref{sec:ftir}). Figure \ref{fig:3microns} also
illustrates that the spectrum of $^{12}$CH$_{3}$D is a scrambled
version of $^{12}$CH$_{4}$, but with shallower features.

In overview, the absorption spectra of $^{14}$NH$_{3}$ and
$^{12}$CH$_{4}$ can be simulated to a high degree of realism by
using the line lists available in the HITRAN 2008 database and a
basic ray-tracing software. Concerning the ray-tracing software, we
explored several options and found that the Reference Forward Model
package (RFM) \footnote{Maintained by Anu Dudhia,
\texttt{http://www.atm.ox.ac.uk/RFM/}} provided the most
straighforward and simplest approach to obtain the desired
synthetic spectra. As a cross-check, We compared our simulated
spectra with those computed with
\textit{SpectralCalc}\footnote{Available at :
\texttt{http://www.spectralcalc.com/}}, obtaining perfect agreement
among the two. As discussed before, the absorption spectra of
$^{13}$CH$_{4}$ is obtained to the required level of realism by
applying a multiplicative factor to the wavelengths of the
$^{12}$CH$_{4}$ spectrum. As will be shown in Sec. \ref{sec:ftir},
this approach worked very well in the estimation of the optimal
pressure for the $^{13}$CH$_{4}$ cell. Because the absorption
spectrum of $^{12}$CH$_{3}$D is very different from
$^{12}$CH$_{4}$, we could not perform the same optimization
analysis. Since the $^{12}$CH$_{3}$D lines at 3.5 microns are
shallower than those from $^{12}$CH$_{4}$, we tentatively filled
this cells with a pressure slighty higer than the optimal one found
for the $^{13}$CH$_{4}$ cell. As shown later, such pressure turned
to be insufficient to produce deep-enough lines. Now that the
spectrum of $^{12}$CH$_{3}$D is known, we will be able to use it to
optimize future cells. Hereafter, all the optimization details
refer only to $^{14}$NH$_{3}$ and $^{13}$CH$_{4}$.

\subsection{Model Configuration Setup}

For the purposes of our optimization models, the length of all the
cells is fixed at 10 cm.  We also assume an ideal spectrograph
continuously covering an interval of 200 nm at the K band. While an
instrument with such capabilities does not yet exist, a comparable
wavelength coverage should be within the reach of the proposed NIR
spectrographs (e.g., i-Shell on NASA's IRTF, and upgraded versions
of NIRSPEC/Keck and CRIRES/VLT). Such wavelength range is also
representative of the interval where the cells under discussion
(ammonia and methane) show more absorption features in the K band.
Note that the cells only provide good wavelength calibration on the
spectral region well covered by them so using the full K band to
estimate the performance of each cell would seriously underestimate
the stellar contribution to the error budget. The central
wavelength of the interval is also obtained during the optimization
process. Four spectral resolutions are also tested: 30 000, 50 000,
70 000 and 10$^5$. These resolutions roughly match the range of
available (or planned) nIR spectrographs. To make a fair
comparison, we will assume that the number of collected photons is
the same in all the setups. In other words, when the resolution is
higher, the signal will be spread on a larger number of resolution
elements and, therefore, the S/N per resolution element will be
reduced. The resulting $\sigma_{V}$ for various setups and gas cell
configurations are given in Figure \ref{fig:results}. In addition to
the cell parameters, the central wavelength  $\lambda_c$ of the 200
nm window is also optimized. The atmospheric K-band window is
surrounded by strong and very variable water absorption features
that we want to avoid as much as possible. Neither gas shows enough
absorption features below 2100 nm to be useful for wavelength
calibration. As a result the useful wavelength range we test is
between 2100 and 2450 nm. For the stellar spectra, we have used
those provided by the PHOENIX  \footnote{For more information, see
\texttt{http://www.hs.uni-hamburg.de/EN/For/ThA/phoenix/index.html}}
group \citep{hauschildt:1999}. Solar metallicity and a $log g =
5.0$ have been assumed in all cases.

\section{Model Results}

For $^{14}$NH$_{3}$, $\sigma_V$ is computed on pressures ranging
from 25 to 250 mbar in steps of 25 mbars. For $^{13}$CH$_{4}$, the
tested pressures go from 50  to 500 mbars in steps of 50 mbars.
Figure \ref{fig:results} and Table \ref{tab:results} summarize our
results as follows :

\begin{itemize}

\item The optimal gas pressure depends on the spectral resolution.
When the pressure is too low, the gas column density is also low and
the lines are shallower. When the pressure is high, the lines are
deeper due to the increased column density, but they get broader due
to presure broadening effects. The optimal gas presures for the
proposed setups are given in Table \ref{tab:results}.

\begin{deluxetable}{lcccccccc}
\tablecaption{Optimal setup for the various test setups. We use
a stellar atmosphere model for a T=3000 K star, and a cell
Length of 10 cm.\label{tab:results}
}
\tablehead{
\colhead{} &
\colhead{Spectral} & 
\colhead{S/N/$\delta\lambda$} & 
\colhead{N$_{\delta\lambda}$} & 
\colhead{$\lambda_{c}$} & 
\colhead{$\sigma_{\rm cell}$} & 
\colhead{$\sigma_*$} &
\colhead{$\sigma_{V}$} &
\colhead{Pressure} \\
\colhead{Gas} &
\colhead{Resolution} & 
\colhead{} & 
\colhead{} & 
\colhead{[nm]} & 
\colhead{[m s$^{-1}$]} & 
\colhead{[m s$^{-1}$]} &
\colhead{[m s$^{-1}$]} &
\colhead{[mb]} 
}
\startdata
$^{13}$CH$_{4}$ \\ 
       & 30 000   & 182 & 2370 & 2370 &  8.9 & 12.3 & 15.1 & 400 \\ 
       & 50 000   & 141 & 4300 & 2360 &  6.5 &  8.0 & 10.4 & 300 \\ 
       & 70 000   & 119 & 6021 & 2360 &  4.8 &  6.6 &  8.2 & 200 \\ 
       & 100 000  & 100 & 8602 & 2362 &  3.8 &  5.7 &  6.9 & 150 \\ 
$^{14}$NH$_{3}$ \\
       & 30 000   & 182 & 2370 & 2298 & 16.7 & 14.0 & 21.8 & 150 \\ 
       & 50 000   & 141 & 4300 & 2292 & 12.4 &  9.3 & 15.5 & 100 \\ 
       & 70 000   & 119 & 6021 & 2298 & 10.6 &  7.4 & 12.9 &  75 \\ 
       & 100 000  & 100 & 8602 & 2298 &  8.7 &  6.5 & 10.9 &  50 \\
\enddata
\end{deluxetable}

%\begin{wrapfigure}{R}{1.0\textwidth}
\begin{figure}[tb]
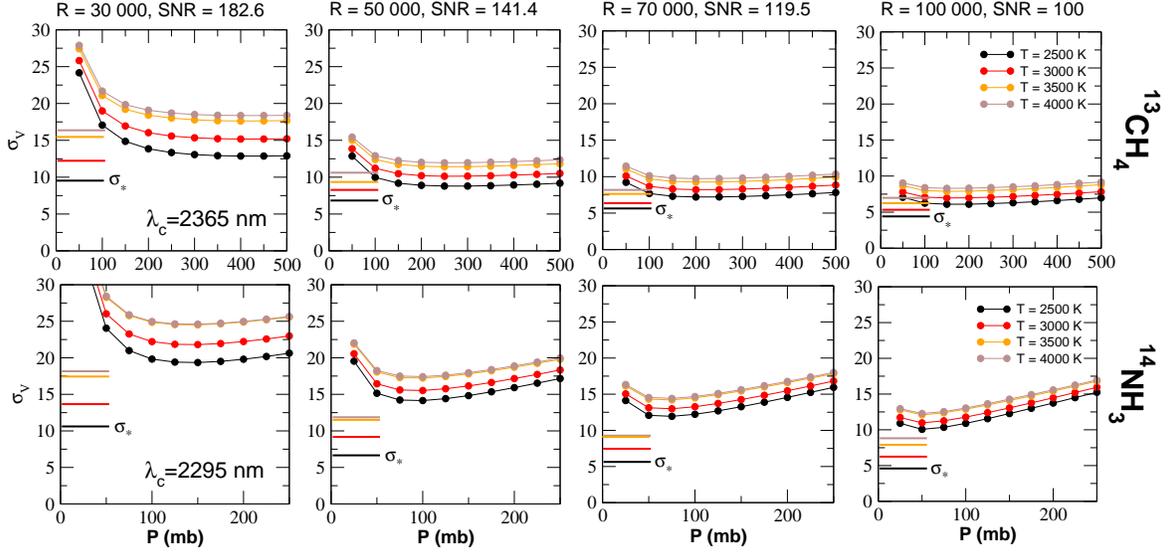

    \begin{center}
        \includegraphics[angle=0, width=6.05in, clip]{f2a.eps}
        \includegraphics[angle=0, width=6.0in, clip]{f2b.eps}
%        \plotone{f2a.eps}
%        \plotone{f2b.eps}
    \end{center}
    
\caption{Photon noise radial velocity precision $\sigma_V$ as a
function of gas cell pressure for different stellar atmosphere models
(colored dots). Because the spectra of stars depends on the effective
temperature, $\sigma_V$ will also depend on that. Top panels contain
the results for the $^{13}$CH$_{4}$ cell and the results for $^{14}$NH$_{3}$ are given in
the bottom panels. The contribution of the stellar spectra (small
horizontal lines on the bottom left of each plot) are also
illustrated. Note that the stellar contribution is different for each
cell. Because the cells have a higher density of lines on slightly
different wavelength ranges, the optimal central wavelength is
different for each cell. Such optimal wavelength also
weakly depends on the resolution and the gas pressure. The average
values of the optimal wavelengths are also given for each setup. In
all the cases, $^{13}$CH$_{4}$ nominally outperforms $^{14}$NH$_{3}$.}
\label{fig:results}

\end{figure}

\item Even accounting for the reduction in the S/N per resolution
element, the higher spectral resolution $R$ always provides higher
RV precision. The spectrum of the star and the cell are
sufficiently resolved at R=100 000, so higher resolutions do not
provide a significant improvement \citep{mahadevan:2009}. Note that
we are not discussing the slit size required to achieve each
resolution. If a fraction of light $L$ is lost due to bad seeing
($>$1.0'') and/or narrow slit (eg. 0.2''), $\sigma_V$ has to be
divided by $\sqrt{L}$. The photon-collection efficiency (also
called \textit{throughput}) for a given mode heavily depends on
engineering details of each spectrograph (slit size, use of an
image slicer, adaptive optics, etc.). The loss of throughput can
easily counter the gain in precision from the higher spectral
resolution. Therefore, one should first compare the relative
performance of the available modes before building a cell optimized
for the highest spectral resolution available. As a general
approach, we suggest first determining the desired $R$ using
synthetic stellar spectra and all the information available on the
available observing modes. Only then one should proceed to the gas
cell optimization for a given R.

\item Both the stellar and the cell spectra contribute
significantly to $\sigma_{V}$. In the case of $^{14}$NH$_{3}$, the
absorption cell dominates $\sigma_{V}$. As a consequence, the RV
measurements will be calibration-noise-limited. On the other
hand, the contribution of $^{13}$CH$_{4}$ to the overall precision
is always smaller than the stellar contribution, which makes it
more attractive than $^{14}$NH$_{3}$. 

\item $^{14}$NH$_{3}$ and $^{13}$CH$_{4}$ cover a different range of
wavelengths. Therefore, the optimal central wavelength $\lambda_c$
strongly depends on the gas used. To a much lower degree,
$\lambda_c$ also depends on the spectral resolution and pressure;
therefore, the optimal wavelength is always within 5 nm to the values
given in Figure \ref{fig:results}. Also, $^{13}$CH$_{4}$ covers the
stellar CO absorption bands better than $^{14}$NH$_{3}$, minimizing
$\sigma_*$ and therefore further improving the maximum achievable
precision. 

\end{itemize}

This latter two items justify the effort of developing a
methane-based absorption cell for future high-precision RV
measurements in the NIR. As we mentioned before, each resolution
element should contain (at least) 2 or more pixels to properly
sample the intrumental profile. If the same S/N can be achieved per
pixel instead, then the maximal RV precisions $\sigma_V$ listed in
Table \ref{tab:results} have to be divided by the square-root of
the number of pixels per resolution element. That is, with the best
setup we tested (R = 100 000, $^{13}$CH$_{4}$, $\sigma_V$ = 6.9 m
s$^{-1}$), a star with T = 3000 K and 2 pixels per resolution
element with a S/N = 100 per pixel, one should be able to achieve
RV precisions better than 5 m s$^{-1}$.

\section{Optimal absorption cells for IRTF/CSHELL}

We started a pilot program to test the absorption cells at the
CSHELL spectrograph installed at the NASA/IRTF telescope
\citep[Mauna Kea,Hawaii; ][]{tokunaga:1990,greene:1993}. The design
parameters and the optimal setup found for this spectrograph are
given in Table \ref{tab:cshell}. In addition to testing their
performance, we were also interested in the practical issues
involved in the construction, installation and operation of such
cells. With this in mind, we built two methane-based cells
containing $^{13}$CH$_{4}$ and $^{12}$CH$_{3}$D, as well as a
$^{14}$NH$_{3}$ cell for comparison purposes. The construction
details and the final high-resolution FTIR of the cells are given
in Section \ref{sec:ftir}. For this experiment, the central
wavelength was not optimized. To simplify the analysis required to
obtain precise RV measurements, we chose $\lambda_{c}$ = 2312.5 nm
which is centered in a small window that is relatively free of
telluric methane features (see  Section \ref{sec:ftir}).

\begin{deluxetable}{lcccccccc}
\tablecaption{Design parameters and optimal setup for
IRTF/CSHELL. The stellar model used in the optimization process
has T = 3500 K, log g = 4.5 and solar metalicity
\label{tab:cshell}}
\tablehead{
\colhead{Parameter} &
\colhead{Value}
}
\startdata
Spectral resolution     & 46 000 \\
Cell length             & 12.5 cm  \\
Cell temperature        & 283 K  \\
Wavelength interval               \\
at K band               & $\sim$ 5 nm \\
N$_{pix}$               & 256    \\
Optimal $^{13}$CH$_{4}$ pressure & 275 mb \\ 
Optimal $^{14}$NH$_{3}$ pressure    & 70 mb \\
Selected $\lambda_{c}$  & 2312.5 nm \\
$\sigma_{V}$ (S/N=100/$\delta \lambda$) & $\sim$ 50 m/s$^{-1}$ \\
$\sigma_{V}$ (S/N=100/pix)
                        & $\sim$ 35 m/s$^{-1}$ \\
\enddata
\end{deluxetable}

\subsection{Construction of the Cells}

Here, we summarize the practical details involved in the construction
of the three cells filled with $^{13}$CH$_{4}$, $^{12}$CH$_{3}$D,
and $^{14}$NH$_{3}$. In all cases, the construction of the cells is
very affordable.

Based on the limitations imposed by the existing extra space in
CSHELL, the bodies of the cells were made using a Pyrex tube 12.5
cm in length and 5.1 cm in diameter. Both ends of this tube were
capped with low-OH quartz windows with excellent transmission in
the NIR. The windows were then coated on the outside with an
infrared antireflective coating to further improve transmission. 
We did not coat the inside of the windows due to potential chemical
reactions of the gas with the coating degrading throughput. Using
the laboratory FTIR spectra of all the cells, we found that the
overall transmission on the continuum (including the gases) is
typically above 80\%.  Each window is a wedge with an angle of
1.6$^{\circ}$ oriented 180$^{\circ}$ relative to the other window
on the cell, corresponding to a 2 mm rise in window thickness over
7 cm, with a minimum window thickness at one end of 1 mm. This
prevents ghost images from appearing on the stellar spectra
which arise from multiple reflections off the spectrograph's
optics. The windows were sealed using Varian Torr Seal, a
sealant specifically designed to connect glass and maintain a
vacuum for a duration longer than a decade in nominal conditions.

Pure anhydrous $^{14}$NH$_3$ ammonia was purchased from
Matheson-TriGas ($>$99.9\% quoted purity). The $^{13}$CH$_{4}$
isotopologue was purchased from Sigma-Aldrich (99\% quoted purity).
The more exotic $^{12}$CH$_{3}$D isotopologue was obtained from
Cambridge Isotope Laboratories, Inc. (98\% quoted purity). All
three cells were evacuated using a standard vacuum pump, and then
filled with the gases. As shown in Table \ref{tab:cshell},
$^{14}$NH$_{3}$ and $^{13}$CH$_{4}$ cells were found to operate
optimally at $\sim$75 mbar and 275 mbar respectively (as a
reference, 1013.25 mbar is 1 atm or 1.013$^5$ Pa). Since we had no
means to predict the optimal pressure for $^{12}$CH$_{3}$D at the K
band, we filled the $^{12}$CH$_{3}$D cell at a higher pressure
($\sim$ 345 mbar), waiting for the FTIR spectra to evaluate if the
cell would be suitable for use at CSHELL. The cells were filled at
such pressures at 20$^o$ C by means of a small inlet in the side of
the pyrex tube. In order to seal this inlet, the gases were then
condensed to a liquid via immersion in an ice bath. This allowed
the Pyrex inlet to be heated to $\sim$ 1100 K without raising the
interior gas temperature and pressure. If the gases had been left
in the gas phase at 20$^o$ C, the internal pressure would have exceeded
1 atmosphere and the gas would have burst out of the hot malleable
inlet. The first tests filling our gas cells with ammonia gas
resulted in this outcome. The end result of the sealing process
allowed for the creation of an essentially permanent Pyrex seal to
the inlet such that the main body tube is a single piece of Pyrex.

\begin{figure}[tb]
  \begin{center}
   \includegraphics[width=0.8\textwidth, clip]{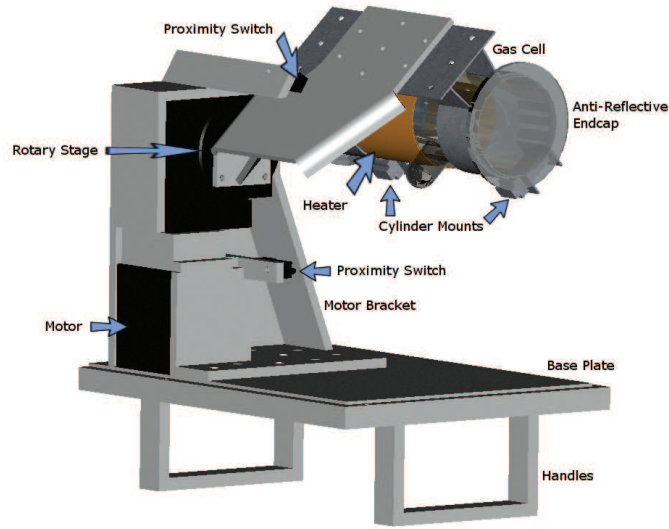}
  \end{center}
 
  \caption{A rendering of the CAD design of the gas cell, rotary
stage, and mounting mechanism. Note the heating element and the
thickness of the aluminum braces (shown in gray) to maintain
rigidity and prevent rotation of the cell for consistent placement
in the beam. Wires and bolts are omitted for visual
clarity.}\label{fig:cad}

\end{figure}

All three cells are interchangeable and can easily be substituted as
needed by removing the current cell and placing one of the others in
the mount. The active cell is mounted inside a calibration box with
an aluminum housing, which in turn sits in front of the CSHELL
spectrograph entrance window.  The gas cell tube is attached to a
rotary stage by aluminum braces which allows the cell to be moved in
and out of the telescope's beam by remote operation (see Fig.
\ref{fig:cad}). The remote operation is an essential design feature
for ease of use and efficiency of observations, given that CSHELL
mounts at the telescope Cassegrain focus. The cell sits in the
converging f/38 beam prior to the beam's entrance into the
spectrograph's entrance window. Thanks to this, the cell absorption
spectrum is imprinted on the stellar light before the light goes
into the spectrograph optics. The available physical space
limitations for the cell ($\sim$ 15 $\times$15 $\times$18 cm$^3$)
placed severe design constraints on the size of the cell and the
motor mechanism to move the cell in and out of the telescope beam.
The length of the cell is limited on one end by the entrance window
to the calibration unit, and on the other end by the descending fold
mirror for the calibration lamps.

% Figure 2 - cad design
%\begin{wrapfigure}{R}{0.5\textwidth}

\begin{figure}[tb]
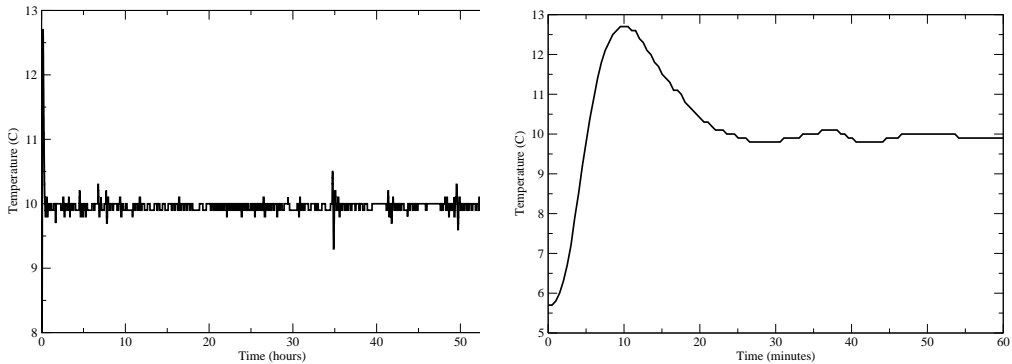

  \begin{center}
    \includegraphics[width=0.4\textwidth]{f4a.eps}    
    \includegraphics[width=0.4\textwidth]{f4b.eps}
      \end{center}
  \caption{Left: Gas cell temperature stability as a function of
hours during first light.  Spikes in the temperature can be seen of
up to 0.5 C, and are due to observations of very bright standards
heating the gas cell.  Right:  Gas cell temperature as a function of
minutes during first hour of turning heater on.  The gas cell
stabilizes at the desired temperature within 1 hour.
\label{fig:temperature}
}
\end{figure}

The IRTF telescope dome experiences ambient temperature ranges of
276 to 284 K. In order to mitigate velocity calibration errors due
to temperature changes of the cell's gas, its temperature is
stabilized with a small silicon heater. For consistency, it is
heated to 283 K (10$^o$ C), at the high end of dome temperatures
experienced over the past year. The cell has an RTD sensor attached
giving temperature feedback to a temperature controller, which is
expected to maintain the temperature to within $\pm$0.1 K. The
temperature controller can be set and logged remotely to ensure
stability. This should result in temperature-induced errors well
below 1 m s$^{-1}$ \citep{bean:2010a}.  As a curiosity, we found
that bright telluric standard stars (e.g. Sirius) heat the cell by
up to $\sim$0.5$^{o}$ C before the temperature controller
re-establishes an equilibrium gas cell temperature (Figure
\ref{fig:temperature}).

% Figure XX - gas cell thermal stability
%\begin{wrapfigure}{R}{1\textwidth}

\subsection{FTIR spectra of $^{14}$NH$_3$, $^{13}$CH$_4$ and CH$_3$D}\label{sec:ftir}

We have obtained laboratory measurements of the three cells'
spectra at the Jet Propulsion Laboratory using a Bruker IFS 125/HR
spectrometer, for which the instrumental setup can be found
elsewhere \citep[e.g.,][]{sung:2008, sung:2012}. The FTIR spectra
were taken at a resolution $R \approx 700 000$ at 2.0 $\mu$m and
298 K. This resolution is much higher than CSHELL's resolution of
$\sim$46 000, and allows for very precise resolution of the
individual spectral absorption features of all three gases. In
order to ensure that we had complete coverage of the infrared bands
we intended to use, a full scan from 1 to 5~$\mu$m wavelengths was
performed.  The FTIR system at JPL is equipped with a
temperature-stabilized He-Neon laser has enabled a frequency
precision better than 0.0001 cm$^{-1}$ in the scanned region, where
the units of frequency are wavenumbers per unit of length. The
internal frequency accuracy is better than 0.01 cm$^{-1}$, that
would correspond to a Doppler offset of 740 m s$^{-1}$ at the K
band. Because precision radial velocity measurements are always
relative, extreme absolute accuracy (as opposed to precision) is
not required for this experiment. If the spectra we provide need to
be used for accurate frequency work, the $^{14}$NH$_{3}$ and
$^{13}$CH$_{4}$ wavelengths can be refined to match the FTIR
spectra to the predicted line positions from HITRAN around 3.0
microns where all the species have abundant (and well-documented)
spectral features.

\begin{figure}[tb]
  \begin{center}
    \includegraphics[width=0.80\textwidth]{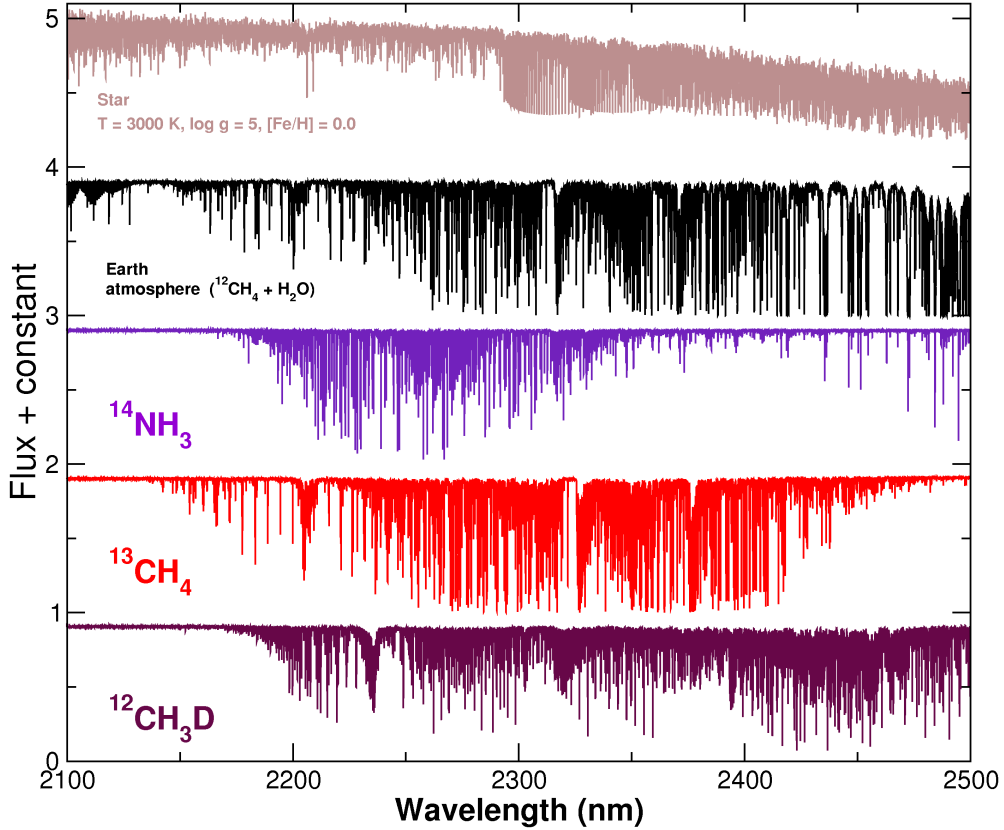}    
      \end{center}
      
  \caption{Relevant spectral features in the K band. From top to
  bottom : synthetic spectrum of an M5V star, absorption spectrum of
  the Earth's atmosphere, and laboratory obtained spectra of our cells :
  $^{14}$NH$_{3}$, $^{13}$CH$_{4}$ and $^{12}$CH$_{3}$D (R$\sim$ 700 000). \label{fig:ftir}}

\end{figure}

We scanned the cell while the FTS was pumped down to 95 to prevent
leaks of ammonia which is known to be sticky and notorious in
leaving permanent residues on the optical surfaces. This pressure
is slightly higher than the cell pressure, minimizing the risk of
leaks. For the methane isotopologues, the FTS was evacuated to
better than 1 mbar in pressure. For all three cases, we obtained
the spectra without the cell at a pressure similar to that of each
cell. In this way, the unwanted atmospheric residual features could
be canceled out by dividing the cell spectra by their \textit{no
cell} counterparts. The normalized K-band FTIR spectra of the cells
are shown in Figure~\ref{fig:ftir}. A synthetic spectrum of a
T=3000~K star ($\sim$ M5V) is shown on the top for comparison. The
second spectrum in Figure \ref{fig:ftir} is a sample of the Earth's
atmosphere absorption along the K band. Compared to the spectra of
$^{13}$CH$_{4}$ (foruth row), it is easy to identify the features
due to telluric methane (eg. look for the gap at 2313 nm, that
appears at 2321 nm in the $^{13}$CH$_{4}$ spectrum). Water vapor is
a major contributor to the telluric absorption beyond 2400 nm.
The $^{14}$NH$_{3}$ cell is in remarkable agreement with the one
obtained by our synthetic spectra generator. As expected, the
$^{13}$CH$_{4}$ spectrum looks very similar to the one from
$^{12}$CH$_{4}$, which validates our optimization procedure. The
$^{12}$CH$_{3}$D spectrum contains a very high density of shallower
lines. Even though $^{12}$CH$_{3}$D has a higher density of lines,
it is not usable for CSHELL because at $R$ = 46 000, those lines
are not deep enough to be competitive against $^{14}$NH$_{3}$ or
$^{13}$CH$_{4}$. Still, our laboratory-obtained spectra of
$^{12}$CH$_{3}$D can now be used to design optimal cells on other
spectrographs. Even though the atmospheric contamination is less
significant around 2150 nm, neither the gas cell nor the stellar
spectra contain many strong absorption features around that
wavelength.

% Figure XX - gas cell thermal stability
%\begin{wrapfigure}{R}{1\textwidth}

Our FTIR spectra were obtained at a temperature of 298 K, whereas
we are operating the gas cell on the IRTF telescope at a
temperature of 283 K. Simulated spectra indicate that the
difference in temperature introduces a small systematic offset of
the order of 1 m/s or less. While this should not affect our
relative RV measurements, the FTIR spectrum of prospective cells
should be obtained at the same telescope operation temperature to
guarantee that the forward modeling of the observed spectra is as
accurate as possible.

\subsection{H Band}

\begin{figure}[tb]
  \begin{center}
    \includegraphics[width=0.80\textwidth]{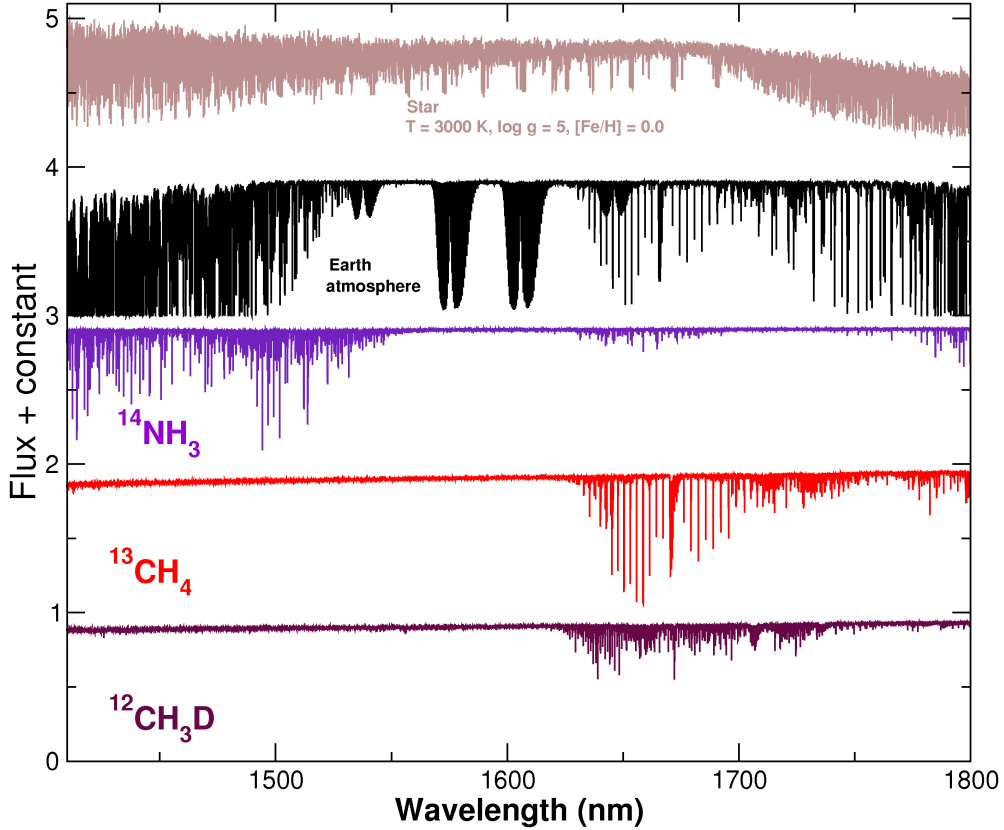}    
      \end{center}
      
  \caption{Relevant spectral features on the H band. From top to
  bottom : synthetic spectrum of an M5V star, absorption spectrum of
  the Earth's atmosphere, and laboratory obtained spectra of our cells
  : $^{14}$NH$_{3}$, $^{13}$CH$_{4}$ and $^{12}$CH$_{3}$D (R$\sim$ 700 000). \label{fig:hband}}

\end{figure}

We show the absorption spectra of our three cells in the H band
(see Fig. \ref{fig:hband}). Similar to K, the H band is surrounded
by very variable water vapor features from the Earth's atmosphere
that should be avoided. The central part of the H band is dominated
by two prominent bands of CO$_2$ that are known to be quite stable
and have been used to reach precision radial velocities at the
level of 10 m s$^{-1}$ \citep{figueira:2010}. Both methane
isotopologues show abundant lines on the redder part of the band,
while ammonia has a very promising band on the bluer part. Even
though the ammonia absorption is quite prominent, such band is not
listed in HITRAN 2008 so it was a surprise to find it there. Since
the gases were not optimized for H band work, the absorption lines
of all three cells are too shallow to produce competitive results
compared with the K band. Testing the H band would require the
construction of additional cells and was beyond the scope of our
limited budget for this initial study. There are other gases and
isotopologues that provide useful absorption features around 1500
nm. Some of the most promising ones have already been discussed by
\citet{mahadevan:2009} and \citet{valdivielso:2010}. From our
obtained spectra, a higher-presure cell combining $^{14}$NH$_{3}$
and $^{13}$CH$_{4}$ would seem to be a good choice to cover a good
fraction of the H band. Because the stellar spectra of cool dwarfs
have fewer features than other IR bands \citep{reiners:2010}, and
because there are other studies focused on this wavelength range,
we do not discuss the H band further.

\subsection{First Light} 

\begin{figure}[tb]
  \begin{center}
    \includegraphics[width=0.80\textwidth,clip]{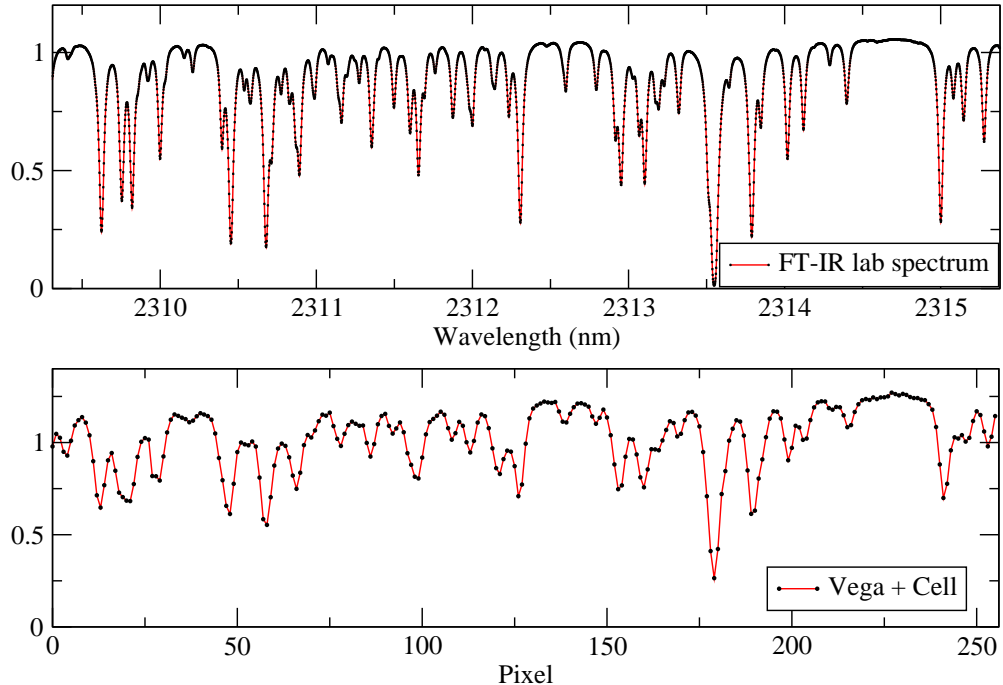}
  \end{center}

  \caption{Laboratory FTIR spectrum (top) and observed spectrum
(bottom) at IRTF of the $^{13}$CH$_{4}$ cell on the K band window
used for CSHELL observations. CSHELL is not cross dispersed and only
covers $\sim$ 6 nm at the K band. The only remarkable  telluric
feature in this window is at pixel 120 and corresponds to telluric
methane absorption. The resolution of the FTIR spectrum is 700 000.
The lines in the observed spectrum look shallower due to the lower
resolution of the spectrograph (R=44 000).}

\label{fig:irtf_vega} 
\end{figure}

\begin{figure}[tb]
  \begin{center}
    \includegraphics[width=0.80\textwidth, clip]{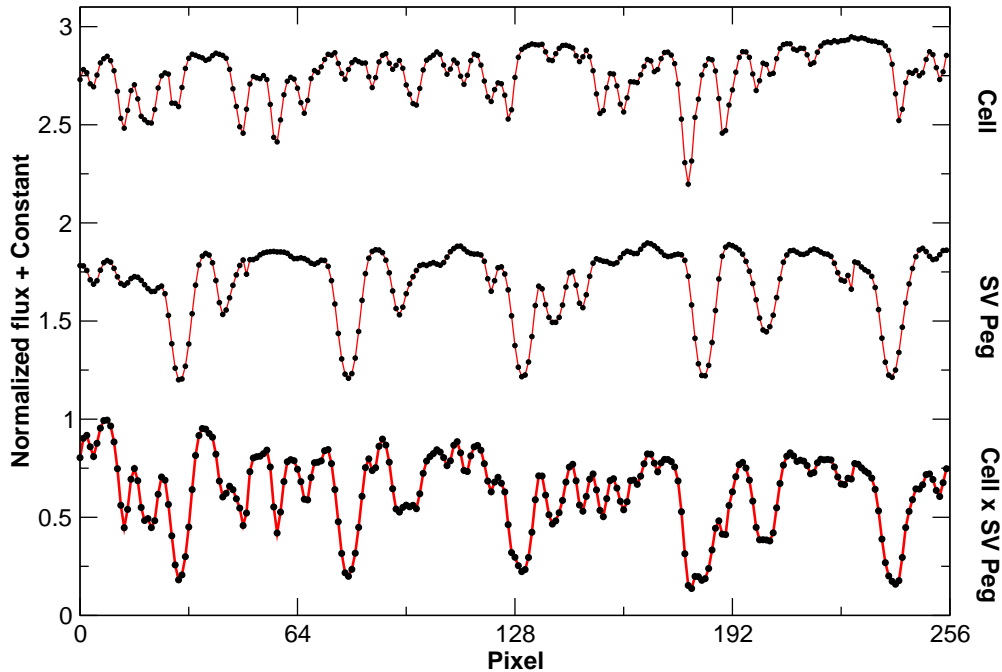}
  \end{center}
  
  \caption{Top. Obtained spectrum of a telluric standard (Vega)
observed through the $^{13}$CH$_4$ absorption cell.  Middle,
observed spectrum of the M7 giant star SV Peg observed without the
cell. Bottom, Observed spectrum of SV Peg through the cell. }

\label{fig:irtf_svpeg}  
 \end{figure}

The $^{13}$CH$_{4}$ cell was successfully integrated in the CSHELL
spectrograph, and first light was achieved on 15 September 2010. We
choose a window centered at $\lambda=2312.5$ nm, because it is
almost free of telluric features. The laboratory spectrum of the
cell compared with the obtained spectrum of a telluric standard
(Vega) through the cell at the telescope is shown in Figure
\ref{fig:irtf_vega}. The spectra confirm the presence of the
methane in the cell, and a wavelength window relatively free of
telluric features. The observed spectra are in excellent agreement
with our expectations. We measure an effective FWHM $\delta \lambda
\sim$ 1.8 pixels in the wavelength direction.  The only prominent
telluric line is present at pixel \#120. Spectra are extracted from
the raw FITS image using a custom pipeline to perform a sum of
counts over the spatial direction as a function of wavelength. The
exposures were taken at two nodding positions separated by 10
$^{\prime\prime}$, which were then subtracted to remove the sky
contribution. The raw frames were also cleaned for hot pixels, dead
pixels, and cosmic-ray events. Several hundred high S/N spectra
(S/N $\sim$ 150, one spectra every 30 seconds) of the supergiant
star SV Peg (M7, K mag = -0.4) were also obtained in the first two
nights, confirming that both the absorption cell features and the
stellar spectrum had abundant lines in the selected wavelength
range (see Figure \ref{fig:irtf_svpeg}). 

\begin{figure}[tb]
  \begin{center}
    \includegraphics[width=0.50\textwidth, clip]{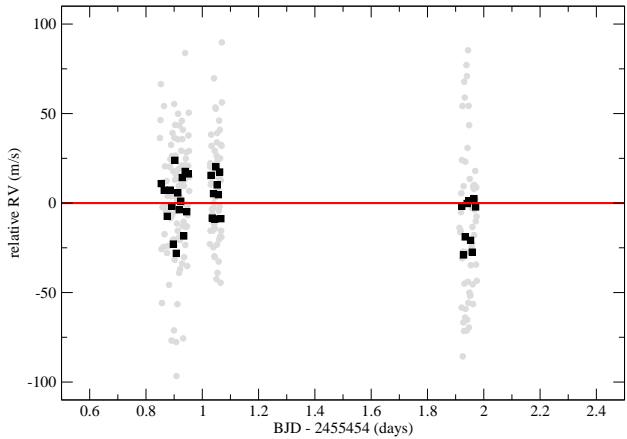}
  \end{center}
  
  \caption{Relative radial velocity measurements of the super
  giant star SV Peg (K = -0.4) obtained during the first two
  nights. Each spectra has a formal S/N of about 150. Grey points
  correspond to  individual observations (RMS $\sim 35$ m
  s$^{-1}$). Black points correspond to averaged blocks of 6
  observations. The measured RMS of the averaged blocks ($\sim
  14.8$ m s$^{-1}$) is almost identical from what would be
  expected from Gaussian white noise ($35/\sqrt{6} = 14.2$ m
  s$^{-1}$). SV Peg is a supergiant star and, as such, should show
  some RV variability in time-scales of a few days and may explain
  the few m s$^{-1}$ offset between the mean of the two nights.}

\label{fig:svpeg_rv}  
 \end{figure}

A preliminary version of our RV extraction pipeline indicates that
a precision of $\sim$ 35 m s$^{-1}$ can be obtained for each
spectrum of SV Peg (see Fig. \ref{fig:svpeg_rv}). Our RV
determination is based on the forward-modeling technique described
by Butler et al. (1996). In brief, we have developed a custom
MATLAB code for our analysis. We convolve a model LSF with a model
for the intrinsic spectrum of a target star, a model for the
telluric spectrum of Earth's atmosphere \citep{wallace:1996}, the
FTIR spectrum for the gas cell, and a foruth-order polynomial model
for the continuum.  The model LSF consists of a central Gaussian
and $N$ satellite Gaussians with adjustable widths, relative
amplitudes and centroids to attempt to reproduced the observed
variability in the instrumental LSF.  We also allow for a variable
spectrograph resolution and plate scale along the length of the
slit.  The residuals between the model and observations are
iteratively minimized over the multiple free parameters via a
hybrid amoeba simplex algorithm from an initial starting set of
parameters that are constrained via a coarse $\chi^2$
minimization.  We derive a deconvolved template spectrum for the
target star using an iterative procedure adding averaged residuals
from our fits using the above procedure to observations taken
without the gas cell. From the fit parameters, we derive the line
of sight radial velocity to a target star from the relative
wavelength shift between the model for the gas cell and the stellar
template.  Using standard barycentric correction routines
\citep{stumpf:1980}, we correct for barycentric motion of the
observer to arrive at the final radial velocities obtained on the
first two nights of observations as shown in Figure
\ref{fig:svpeg_rv}. SV Peg was chosen to be very bright and with
spectral features in the K band, but it is a M supergiant star and,
as such, RV variability at $>10$ m s$^{-1}$ level is expected in
timescales of a few days. Also, as happens in the optical,
obtaining accurate templates is a major limitation of the
absorption-cell method. To derive more reliable templates, we are
now obtaining very high resolution spectra in the K band using
CRIRES/VLT (R$\sim$110 000). The long-term stability of our setup
will be demonstrated in a forthcoming publication using
observations on known RV stable M dwarfs (e.g., GJ 15A and GJ 293).

%\begin{wrapfigure}{R}{1.0\textwidth}

%\begin{wrapfigure}{R}{1.0\textwidth}

\section{Conclusions and Current work}

We demonstrated a methodology to design optimal gas cells for
precision RV measurements using high-resolution spectrographs. Our
numerical experiments showed that, given a high resolution
spectrograph covering the K band, precisions better than 5 m
s$^{-1}$ level can be achieved for late M dwarfs (T$<$3500 K),
enabling the detection of terrestrial planets orbiting in their
habitable zones. Additionally, precision RVs of earlier type G0 to
mid-K dwarfs could also be obtained, enabling planet-search
programs around young active stars. We constructed two such methane
isotopologue cells and presented FTIR spectra useful for future RV
applications and isotopic abundance determination on solar-system
studies. We commissioned the $^{13}$CH$_{4}$ gas cell on
CSHELL/IRTF, and provided optimal parameters for future
spectrographs with the ability to operate in the K band. Even if
the final $^{12}$CH$_{3}$D cell was suboptimal for work in the K
band, we found that this isotopologue of Methane has an even higher
density of lines. Thanks to the obtained FTIR spectra, a cell with
$^{12}$CH$_{3}$D can also be now optimized. A similar increase in
line density would be expected from a deuterated isotopologue of
ammonia (e.g., $^{14}$NH$_2$D). Unfortunately, no comprehensive
line lists exist for such species. We plan to better characterize
some of these isotopologues in the future using laboratory FTIR
spectroscopy and custom-made cells.

The absorption-cell technique discussed here consists of inserting
the cells directly on the optical path of the starlight. Given a
stabilized spectrograph (e.g., HARPS/ESO), such cells could also be
used as external calibration sources by illuminating them with
white light and producing a `lamp-like' calibration spectrum in
absorption \citep[see ][for a more detailed
discussion]{mahadevan:2009}. However, stabilized spectrographs are
more expensive to build, tend to be less versatile and none is
currently available to work in the NIR. Alternative methods to
provide external wavelength calibration have been proposed (e.g.,
frequency comb and/or a stabilized ethalon), but they will also
work on stabilized spectrographs. The absorption-cell technique
remains as the only option if precision RV measurements are needed
from general-purpose instruments such as NIRSPEC/Keck, CRIRES/VLT
or the planned i-Shell/IRTF.

In October 2010, we started a pilot program to obtain RV
measurements on late-type young dwarfs (K and M). The targeted
radial velocity precision is between 30 to 50 m s$^{-1}$ (depending
on the spectral type). An end-to-end data analysis pipeline is
being developed that applies the forward modeling technique
outlined in \citet{butler:1996}. Stellar spectra templates at
higher resolution (using CRIRES/VLT, R=110 000) are being obtained
to be used in the modeling of the observed spectrum. The first
results of our survey and the long-term stability of the methane
cells will be presented in a forthcoming publication. Preliminary
intranight measurements on the giant star SV Peg (K mag $=$ -0.4)
indicate that a precision down to 35 m s$^{-1}$ (S/N = 150) is
within our reach. Let us note that we are only using 6 nm out of
300 nm avaliable in the K band. Preliminary analysis of the most
recent observing run (August 2011) confirms that the
$^{13}$CH$_{4}$ gas cell pressure and operation remains nominal
after nearly one year of operation on the telescope at the summit
of Mauna Kea, Hawaii. Given the better nominal performance of
methane isotopologues compared to $^{14}$NH$_{3}$, and assuming
that the photon noise is a significant term in the final error
budget, a $^{13}$CH$_{4}$ (or $^{12}$CH$_{3}$D) cell installed on a
CRIRES-like spectrograph and a similar setup as the one used by
\citet{bean:2010b} should lead to a 30-40\% improvement in the
overal Doppler accuracy (3-4 m s$^{-1}$ precision compared to the
current 6 m s$^{-1}$ long term accuracy demonstrated on Barnard
star and Proxima Cen). \citep{bean:2010b} discussed that the likely
limiting factor of the Ammonia-CRIRES program was contamination
by shallow telluric features, both on the observations and during
the stellar template reconstruction. While this is a likely source
of systematic uncertainty, our numerical simulations indicate that
the ammonia cell's contribution to the error budget has almost the
same magnitude as the reported RMS on RV stable stars. We therefore
conclude that, even though telluric contamination is a source of
uncertainty for sure, using a methane cell on CRIRES should lead to
a significant increase in precision. Given that enough space is
left between the telescope and the spectrograph ($\sim 10$ cm),
such cells can be installed at little cost in upgraded versions of
the available NIR instruments (eg. NIRSPEC at Keck, CRIRES/VLT) and
planned instruments with K-band capabilities (e.g., i-Shell at
NASA/IRTF).

\subsection*{Acknowledgements}

Both G. Anglada-Escud\'e and P. Plavchan contributed equally to this
work. G.A. would like to acknowledge the Carnegie Postdoctoral
Fellowship Program and the support provided by the NASA Astrobiology
Institute grant NNA09DA81A. Peter Plavchan would like acknowledge Wes
Traub and Stephen Unwin for funding provided by the JPL Center for
Exoplanet Science and NASA Exoplanet Science Institute.  K. Sung
acknowledges the Planetary Atmospheric Research Program to support
the laboratory spectroscopic calibrations. Part of the research at
the Jet Propulsion Laboratory (JPL) and California Institute of
Technology was performed under contracts with National Aeronautics
and Space Administration. We thank Anu Dudhia for making the RFM code
available to us and his assistance adapting it for gas cell spectral
calculations. The stellar synthetic spectra were graciously provided
by Peter Hauschildt (U. of Hamburg) and the PHOENIX group. We also
thank Linda Brown from JPL's  Laboratory Studies \& Modeling group,
and Pin Chen from JPL's Planetary Chemistry  \& Astrobiology group
for their advice and support using the  FTIR spectrometer. We would
like to thank Paul Butler (Carnegie Institution of Washington) and
Gilian Nave (NIST) for their advice in gas optimization parameters
and mollecular spectroscopy in general. We would like to thank
Stephen Kane (NExScI), Kaspar von Braun (NExScI), and Steve Osterman
(U. of Colorado) for their valuable discussions. We also thank John
Rayner, Morgan Bonnet, George Koenig and Alan Tokunaga from
IfA/Hawaii for their support during the CSHELL/IRTF cell design
review, integration and commissioning.  We thank Rick Gerhart
(Caltech), Scot Howell (Mindrum Precision) and Thurston Levy (Glass
Instruments, Inc.) for their work in helping construct and fill the
gas cells, and Joeff Zolkower (Caltech) for mechanical engineering
advise.

%\bibliography{biblio}

\begin{thebibliography}{34}
\expandafter\ifx\csname natexlab\endcsname\relax\def\natexlab#1{#1}\fi

\bibitem[{{Bailey} {et~al.}(2011){Bailey}, {White}, {Blake}, {Charbonneau},
  {Barman}, {Tanner}, \& {Torres}}]{bailey:2011}
{Bailey}, J., {White}, R., {Blake}, C., et al. 2012, accepted in ApJ, --,

\bibitem[{{Bean} {et~al.}(2010{\natexlab{a}}){Bean}, {Seifahrt}, {Hartman},
  {Nilsson}, {Reiners}, {Dreizler}, {Henry}, \& {Wiedemann}}]{bean:2010b}
{Bean}, J.~L., {Seifahrt}, A., {Hartman}, H. et al. 2010{\natexlab{a}}, \apjl,
  711, L19

\bibitem[{{Bean} {et~al.}(2010{\natexlab{b}}){Bean}, {Seifahrt}, {Hartman},
  {Nilsson}, {Wiedemann}, {Reiners}, {Dreizler}, \& {Henry}}]{bean:2010a}
{Bean}, J.~L., {Seifahrt}, A., {Hartman}, H. et al. 2010{\natexlab{b}}, \apj,
  713, 410

\bibitem[{{Boudon} {et~al.}(2009){Boudon}, {Champion}, {Gabard}, {Loete},
  {Coustenis}, {De Bergh}, {B\'ezard}, {Lellouch}, {Drossart}, {Hirtzig},
  {Negrao}, \& {Griffith}}]{titan:2009}
{Boudon}, V., {Champion}, J., {Gabard}, T. et al. 2009, Europhys. News., 40, 17

\bibitem[{{Butler} {et~al.}(1996){Butler}, {Marcy}, {Williams}, {McCarthy},
  {Dosanjh}, \& {Vogt}}]{butler:1996}
{Butler}, R.~P., {Marcy}, G.~W., {Williams}, E. et al. 1996, \pasp, 108, 500

\bibitem[{{Charbonneau} {et~al.}(2009){Charbonneau}, {Berta}, {Irwin}, {Burke},
  {Nutzman}, {Buchhave}, {Lovis}, {Bonfils}, {Latham}, {Udry}, {Murray-Clay},
  {Holman}, {Falco}, {Winn}, {Queloz}, {Pepe}, {Mayor}, {Delfosse}, \&
  {Forveille}}]{charbonneau:2009}
{Charbonneau}, D., {Berta}, Z.~K., {Irwin}, J., et al. 2009, \nat, 462, 891

\bibitem[{{Crane} {et~al.}(2010){Crane}, {Shectman}, {Butler}, {Thompson},
  {Birk}, {Jones}, \& {Burley}}]{crane:2010}
{Crane}, J.~D., {Shectman}, S.~A., {Butler}, R.~P., et al. 2010, 
  SPIE Conf. Series, Vol. 7735

\bibitem[{{Endl} {et~al.}(2006){Endl}, {Cochran}, {K{\"u}rster}, {Paulson},
  {Wittenmyer}, {MacQueen}, \& {Tull}}]{endl:2006}
{Endl}, M., {Cochran}, W.~D., {K{\"u}rster}, M., {Paulson}, D.~B. et
al. 2006, \apj, 649, 436

\bibitem[{{Figueira} {et~al.}(2010){Figueira}, {Pepe}, {Melo}, {Santos},
  {Lovis}, {Mayor}, {Queloz}, {Smette}, \& {Udry}}]{figueira:2010}
{Figueira}, P., {Pepe}, F., {Melo}, C.~H.~F., et al. 2010, \aap, 511, A55+

\bibitem[{{Ge} {et~al.}(2002){Ge}, {Erskine}, \& {Rushford}}]{ge:2002}
{Ge}, J., {Erskine}, D.~J., \& {Rushford}, M. 2002, \pasp, 114, 1016

\bibitem[{{Gillon} {et~al.}(2007){Gillon}, {Pont}, {Demory}, {Mallmann},
  {Mayor}, {Mazeh}, {Queloz}, {Shporer}, {Udry}, \& {Vuissoz}}]{gillon:2007}
{Gillon}, M., {Pont}, F., {Demory}, B.-O., et al. 2007,
  \aap, 472, L13

\bibitem[{{Greene} {et~al.}(1993){Greene}, {Tokunaga}, {Toomey}, \&
  {Carr}}]{greene:1993}
{Greene}, T.~P., {Tokunaga}, A.~T., {Toomey}, D.~W., \& {Carr}, J.~B. 1993, in
  SPIE Conf. Series, Vol. 1946, p313--324

\bibitem[{{Hauschildt} {et~al.}(1999){Hauschildt}, {Allard}, \&
  {Baron}}]{hauschildt:1999}
{Hauschildt}, P.~H., {Allard}, F., \& {Baron}, E. 1999, \apj, 512, 377

\bibitem[{{Howard} {et~al.}(2010){Howard}, {Marcy}, {Johnson}, {Fischer},
  {Wright}, {Isaacson}, {Valenti}, {Anderson}, {Lin}, \& {Ida}}]{howard:2010}
{Howard}, A.~W., {Marcy}, G.~W., {Johnson}, J.~A. et al. 2010, Science, 330, 653

\bibitem[{{Huang} {et~al.}(2008){Huang}, {Schwenke}, \& {Lee}}]{nh3:2008}
{Huang}, X., {Schwenke}, D., \& {Lee}, T. 2008, J. Chem. Phys., 129, 214304

\bibitem[{{Johnson} {et~al.}(2007){Johnson}, {Butler}, {Marcy}, {Fischer},
  {Vogt}, {Wright}, \& {Peek}}]{johnson:2007}
{Johnson}, J.~A., {Butler}, R.~P., {Marcy}, G.~W., et al. 2007, \apj, 670, 833

\bibitem[{{Kaeufl} {et~al.}(2004){Kaeufl}, {Ballester}, {Biereichel},
  {Delabre}, {Donaldson}, {Dorn}, {Fedrigo}, {Finger}, {Fischer}, {Franza},
  {Gojak}, {Huster}, {Jung}, {Lizon}, {Mehrgan}, {Meyer}, {Moorwood}, {Pirard},
  {Paufique}, {Pozna}, {Siebenmorgen}, {Silber}, {Stegmeier}, \&
  {Wegerer}}]{crires:2004}
{Kaeufl}, H.-U., {Ballester}, P., {Biereichel}, P.,  et al. 2004, SPIE 
Conf. Series, Vol. 5492, 1218--1227

\bibitem[{{Mahadevan} \& {Ge}(2009)}]{mahadevan:2009}
{Mahadevan}, S. \& {Ge}, J. 2009, \apj, 692, 1590

\bibitem[{{Mayor} {et~al.}(2009){Mayor}, {Bonfils}, {Forveille}, {Delfosse},
  {Udry}, {Bertaux}, {Beust}, {Bouchy}, {Lovis}, {Pepe}, {Perrier}, {Queloz},
  \& {Santos}}]{mayor:2009}
{Mayor}, M., {Bonfils}, X., {Forveille}, T., et al. 2009, \aap, 507, 487

\bibitem[{{McLean} {et~al.}(1998){McLean}, {Becklin}, {Bendiksen}, {Brims},
  {Canfield}, {Figer}, {Graham}, {Hare}, {Lacayanga}, {Larkin}, {Larson},
  {Levenson}, {Magnone}, {Teplitz}, \& {Wong}}]{nirspec:1998}
{McLean}, I.~S., {Becklin}, E.~E., {Bendiksen}, et al. 1998, SPIE
  Conf. Series, Vol. 3354, p566--578

\bibitem[{{Nai-Cheng} {et~al.}(1981){Nai-Cheng}, {Yao-Xiang}, {Yi-Min},
  {Cheng-Yang}, {Xue-Bin}, \& {Chu}}]{methaneref1}
{Nai-Cheng}, S., {Yao-Xiang}, W., {Yi-Min}, S., {Cheng-Yang}, L., {Xue-Bin},
  Z., \& {Chu}, W. 1981, in Precision Measurement and Fundamental Constants,
  ed. {B.~N.~Taylor \& W.~D.~Phillips}, 77--+

\bibitem[{{Reiners} {et~al.}(2010){Reiners}, {Bean}, {Huber}, {Dreizler},
  {Seifahrt}, \& {Czesla}}]{reiners:2010}
{Reiners}, A., {Bean}, J.~L., {Huber}, K.~F., {Dreizler}, S., {Seifahrt}, A.,
  \& {Czesla}, S. 2010, \apj, 710, 432

\bibitem[{{Rivera} {et~al.}(2010){Rivera}, {Laughlin}, {Butler}, {Vogt},
  {Haghighipour}, \& {Meschiari}}]{rivera:2010}
{Rivera}, E.~J., {Laughlin}, G., {Butler}, R.~P., {Vogt}, S.~S.,
  {Haghighipour}, N., \& {Meschiari}, S. 2010, \apj, 719, 890

\bibitem[{Rothman {et~al.}(2009)Rothman, Gordon, Barbe, Benner, Bernath, Birk,
  Boudon, Brown, Campargue, Champion, Chance, Coudert, Dana, Devi, Fally,
  Flaud, Gamache, Goldman, Jacquemart, Kleiner, Lacome, Lafferty, Mandin,
  Massie, Mikhailenko, Miller, Moazzen-Ahmadi, Naumenko, Nikitin, Orphal,
  Perevalov, Perrin, Predoi-Cross, Rinsland, Rotger, Simecková, Smith, Sung,
  Tashkun, Tennyson, Toth, Vandaele, \& Auwera}]{hitran:2008}
Rothman, L., Gordon, I., Barbe, A., et al. 2009, JQSRT, v110, p533

\bibitem[{{Stumpff}(1980)}]{stumpf:1980}
{Stumpff}, P. 1980, \aaps, 41, 1

\bibitem[{{Sung}. {et~al.}(2008){Sung}., {Brown}, {Toth}, \&
  {Crawford}}]{sung:2008}
{Sung}., K., {Brown}, L.~R., {Toth}, R.~A., \& {Crawford}, T.~J. 2008, Canad.
  J. Phys., 87, 469

\bibitem[{{Sung} \& et~al.(2012)}]{sung:2012}
{Sung}, K., {Brown}, L.R., {Huang}, X. et~al. 2012, JQSRT, in print

\bibitem[{{Tokunaga} {et~al.}(1990){Tokunaga}, {Toomey}, {Carr}, {Hall}, \&
  {Epps}}]{tokunaga:1990}
{Tokunaga}, A.~T., {Toomey}, D.~W., {Carr}, J., {Hall}, D.~N.~B., \& {Epps},
  H.~W. 1990, SPIE Conf. Series, Vol. 1235, p 131--143

\bibitem[{{Urban} {et~al.}(1989){Urban}, {Tu}, {Narahari Rao}, \&
  {Guelachvili}}]{urban:1989}
{Urban}, {\v S}., {Tu}, N., {Narahari Rao}, K., \& {Guelachvili}, G. 1989,
  Journal of Molecular Spectroscopy, 133, 312

\bibitem[{{Valdivielso} {et~al.}(2010){Valdivielso}, {Esparza},
  {Mart{\'{\i}}n}, {Maukonen}, \& {Peale}}]{valdivielso:2010}
{Valdivielso}, L., {Esparza}, P., {Mart{\'{\i}}n}, E.~L., {Maukonen}, D., \&
  {Peale}, R.~E. 2010, \apj, 715, 1366

\bibitem[{{Vogt} {et~al.}(2010){Vogt}, {Butler}, {Rivera}, {Haghighipour},
  {Henry}, \& {Williamson}}]{vogt:2010}
{Vogt}, S.~S., {Butler}, R.~P., {Rivera}, E.~J., {Haghighipour}, N., {Henry},
  G.~W., \& {Williamson}, M.~H. 2010, \apj, 723, 954

\bibitem[{{Wallace} {et~al.}(1996){Wallace}, {Livingston}, {Hinkle}, \&
  {Bernath}}]{wallace:1996}
{Wallace}, L., {Livingston}, W., {Hinkle}, K., \& {Bernath}, P. 1996, \apjs,
  106, 165

\bibitem[{Yurchenko {et~al.}(2005)Yurchenko, Zheng, Lin, Jensen, \&
  Thiel}]{nh3:2005}
Yurchenko, S.~N., Zheng, J., Lin, H., Jensen, P., \& Thiel, W. 2005, J. Chem.
  Phys., 123, 134308

\bibitem[{{Zechmeister} {et~al.}(2009){Zechmeister}, {K{\"u}rster}, \&
  {Endl}}]{zec:2009}
{Zechmeister}, M., {K{\"u}rster}, M., \& {Endl}, M. 2009, \aap, 505, 859

\end{thebibliography}

\end{document}